\documentclass[a4paper,twoside,10pt]{letter}
\usepackage{graphicx,saj,multicol,subeqnarray,amsmath}

\def\udc{...}
\begin{document}

\baselineskip=3.1truemm \columnsep=.5truecm
\markboth{\eightrm DIFFICULTIES OF PRESERVING THE LEAP SECOND}
{\eightrm S. {\v S}EGAN, B. {\v S}URLAN, S. VIDOJEVI{\' C}}


\publ



\title{DIFFICULTIES OF PRESERVING THE LEAP SECOND}

\authors{S. {\v S}egan$^{1,}$\footnote[2]{Send comments to Stevo Segan, e-mail: ssegan@matf.bg.ac.yu}, B. {\v S}urlan$^{1}$ and S. Vidojevi{\' c}$^{1}$}

\vskip3mm


\address{$^1$Department of Astronomy, Faculty of Mathematics,
University of Belgrade\break Studentski trg 16, 11000 Belgrade,
Serbia}

\dates{2007}{... }


\summary{In this article we examine the possibility to extend leap
second extrapolation for a near future based on some periodic terms
in the Earth's rotation changes. The IERS data, covering the interval
from 1962.15 to 2006.95, are analyzed.  The difference $\Delta T$ is
extrapolated till to 2035 and compared with the IERS extrapolated
values to the 2012. It can be seen that for the interval from 2006 to
2024 only 1 leap seconds (negative) will be operated.}



\begin{multicols}{2}
{


\section{1. The leap second}

The recommendations of the IAU (International Astronomical Union)
were formalized by resolutions of their Commissions that the name
$UTC$ (abbreviation is compromise between English Coordinated
Universal Time and French  Temps Universel Coordonn\'{e}) was
retained. $UTC$ was recommended as the basis of standard time in all
countries, the time in common (civil) use. The limit of $[UT1 - UTC]$
($UT1$, Universal time) was set at $+0.950$\,s, as this is the
maximum difference that can be accommodated by the code format. The
maximum deviation of $UTl$  from $[ UTC + DUT1]$ (time correction
$DUT1=UT1-UTC$) was set at $+0.100$\,s. In 1974, the CCIR
(Consultative Committee on International Radio, En. or Comit\'{e}
consultatif international pour la radio, Fr.), increased the
tolerance for $[UT1 - UTC]$ from $0.7$\,s to $0.9$\,s. The present
$UTC$ system is defined by ITU-R (International Telecommunication
Union -- Radio) (formerly CCIR) Recommendation ITU--R TF.460--5:

''$UTC$ is the time scale maintained by the BIPM (Bureau
International des Poids et Mesures, Fr. or International Bureau of
Weights and Measures, En.), with assistance from the IERS
(International Earth Rotation and Reference Systems Service), which
forms the basis of a coordinated dissemination of standard
frequencies and time signals. It corresponds exactly in rate with
$TAI$ (International Atomic Time, En., Temps Atomique International,
Fr.) but differs from it by an integral number of seconds. The $UTC$
scale is adjusted by the insertion or deletion of seconds (positive
or negative leap seconds) to ensure approximate agreement with
$UT1$.''

Because the Earth's rotation is gradually slowing down, and in
addition it has both random and periodic fluctuations, it is not a
uniform measure of time. The time difference $\Delta T = [ET - UT1] =
[TT - UT1]$  represents the difference between the uniform scale of
Ephemeris Time ($ET$) or Terrestrial Time ($TT$) and the variable
scale of Universal Time ($UT1$). Before 1955, the values are given by
$\Delta T = [ET- UT1]$ based on observations of the Moon. After 1955,
values are given by $\Delta T = [TT- UT1] = [TAI + 32^s.184 - UT1]$
from measurements by atomic clocks as published by the BIH (Bureau
International de l'Heure, Fr. or International Time Bureau, En.)  and
the BIPM.

According to Stephenson and Morrison (1995), over the past 2700 years
$\Delta T$ can be represented by a parabola of approximately the form
$$\Delta T = (31^s/\mathrm{cy}^2) (T- 1820)^2/(100)^2- 20^s.$$
where $\Delta T$ is expressed in seconds and $T$ is the year. The
minimum is at the year 1820 and passes through 0 at the year 1900.
Actual values of $\Delta T$ based on astronomical data may differ
somewhat from this smoothed fit. For example, the value of $\Delta T$
is $32^s.184$ at 1958.0, the origin of $TAI$. However, no single
parabola can {\bf satisfactorily represent} all the observational
data.

The derivative of $\Delta T$ is
$$\Delta LOD = (0.00 17 \,\mathrm{s}\,\mathrm{d}^{-1}\mathrm{cy}^{-1}) \frac{( T - 1820)}{100}.$$
which represent the change of the length of day ($LOD$) in SI
seconds. Different studies implies different values. The actual value
of the $LOD$ will depart from some long-term trend with short-term
fluctuations (periodicity) between $\pm 3\;$ms on time scale of
decades.

Similarly, the insertion of leap seconds is due to the fact that the
present length of the mean solar day is about $2.5\;$ms longer than a
day of precisely 86400 SI seconds, as a consequence of the long-term
trend, so that the Earth's rotation runs slow with respect to atomic
time.

\section{2. The data and model}

$UTC$ is kept within $0.9$\,s of $UT1$ by the occasional insertion of
a leap second adjustment. When the present $UTC$ system was
established in 1972, the time difference $\Delta T= [TT- UT1] = [TAI+
32^s.184 - UT1]$ was equal to $42.23$\,s. Thus the difference between
$TAI$ and $UT1$ in 1972 was approximately 10\,s. To maintain
continuity with $UT1$, $UTC$ was initially set behind $TAI$ by this
amount. As of  January 1st, 2006 the 23 positive leap seconds have
been added. Thus $UTC$ is presently behind $TAI$ by $33$\,s. Figure 1
illustrates the relationships between $TAI$, $UTC$ and $UT1$ (IERS
data).

%

A least-squares fit of the difference $[UTC-TAI]$ since 1972, shown
in Figure 1, implies a nearly linear increase with a slope of $(2.10
+ 0.05)\;$ms per day. This value represents the average excess in the
length of day during the past three decades and is in approximate
agreement with the value computed on the basis of the long-term
trend.

Recent global weather conditions have contributed to a short-term
change in the length of day. Decade fluctuations due to the
interaction between the Earth's core and mantle and global ocean
circulation may also contribute. The model characterized by
triaxial Earth's structure, its fluid core, visco-elastic mantle
and equilibrium ocean was proposed by the Vondr\'ak (1987).

\noindent\includegraphics[width=8cm,height=6cm]{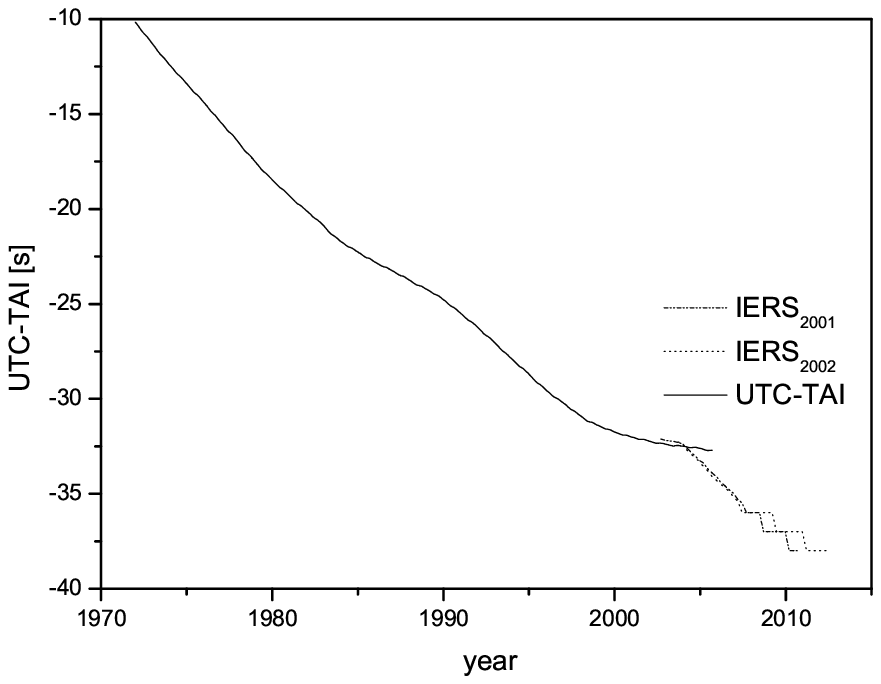}

\noindent{\bf Figure 1.} {\sl $UTC-TAI$ as extrapolated by IERS at
2001 and 2002 compared with observed values of $UTC-TAI$ till
2006.}\vskip.2cm

As a contrary Yoshido and Hamano (1993) have proposed that one of the
main causes of the secular variation of the geomagnetic field is
length-of-day ($LOD$) variation, namely, the variation of the
rotation velocity of the mantle. They developed an analytical model,
in which fluctuations of the rotation velocity of the mantle induce
flow in the outer core through topographic coupling at the
core-mantle boundary (CMB). The flow in turn bends the toroidal field
to produce a poloidal field.

At periods longer than few years, extending to many tens of years,
the so-called {\it decadal variations}, the sources of excitation for
both $LOD$ and PM 
are more enigmatic. The difficulty is that at these periods other
effects may be important, including visco-elastic behavior such as
post-glacial rebound, and exchange of angular momentum with the fluid
core. In particular, it has become common to invoke the core as the
major cause of decadal $LOD$ changes. Although some climatic forcing
of long period $LOD$ has been recognized (Salstein and Rosen, 1986;
Eubanks, 1993), it is uncertain at what time scale air and water
become less important than the core. Unfortunately, the role of the
core remains largely unquantified because it is too remote to be
easily observed. A further difficulty in assessing the air/water role
at long periods is that the torques required to cause decadal $LOD$
variations are utterly insignificant when compared with those applied
by the atmosphere at shorter time scales (Hide and Dickey, 1991).
This means that atmospheric/oceanic torques of geodetic significance
are of second-order importance in general circulation studies.
Quantification of momentum budgets among Earth, air, and water
reservoirs is thus lacking at long periods. The requirements for
progress in this field coincide completely with the central problems
of global climate change.

The five Earth Orientation Parameters (EOP) - two components of polar
motion $x, y$, two components of celestial pole offsets $\Delta \psi,
\Delta \epsilon$, and universal time $UT1$ (that is nothing else but
the angle of Earth's rotation around its spin axis) - are now
analyzed and routinely derived from the observations at several
world's centers, combined and regularly published in IERS bulletins.
The most recent system of constants and algorithms (McCarthy, 1996,
Vondr\'ak, 1998) are used . The past solutions based on optical
astrometry were merged with the combined solutions from the modern
techniques (Vondr\'ak, 2001).

The existing ERP (Earth Orientation Parameters)  series have been
analyzed by many scientists. The most extensive reviews, with
historical meaning, were given in well-known monographs by Munk \&
MacDonald (1960) and Lambeck (1980). Most recent of these analyses is
that by Zharkov et al. (1996). A very detailed review, mostly
concentrated on modern space techniques and discussion of
short-periodic effects, was published by Eubanks (1993).  Moreover,
in our previous paper ({\SH}egan  et al., 2003) we have introduced
the relax period as good explanation of the residuals in the
$(UT1-UTC)_{\mathrm{BLI}}$ data.

The motivation for the leap second, therefore, is due to the fact
that the second as presently defined is "too short" to keep in step
with the Earth. However, had the second been defined {\bf to be
exactly equal} to a mean solar second at the origin of $TAI$ in 1958,
the discrepancy {\bf would not have been removed}; {\bf the
agreement} between the SI second and the mean solar second would have
only been {\bf temporary} and their difference would simply have
become gradually more apparent over this century.

Continuing use of a non-uniform time scale including leap seconds
in the face of these considerations could lead to the necessity to
proliferate an effective method for extrapolation of the future
values of  $\Delta T$.

To try that, according to our analysis of the IERS $\Delta T$ data
from 1962.15  to 2006.95, existence of some periodic terms is
discovered. By using the harmonic analysis 17 components plus a
linear term of equation (1) are determined.
\begin{equation}\label{eq:deltaT}
\begin{split}
M\Delta T_i&= C_1 + C_2\times t_i +\\
 &+\sum_{j=1}^{17} A_j^c \cos\left ( \frac{2\pi}{P_j}\cdot t_i
\right)  + A_j^s \sin\left ( \frac{2\pi}{P_j}\cdot t_i  \right).
\end{split}
\end{equation}
In equation (1) $M\Delta T_i$ is a modified   $\Delta T_i$. In order
to obtain the $\Delta T_i$ values one needs to perform a translation
by a constant value of $32.184 + 10$ seconds:
\begin{equation*}
    \Delta T_i =32.184 + 10 - M\Delta T_i.
\end{equation*}
\vskip3cm

\noindent{\bf Table 1.} {\sl The coefficients for 17 components
obtained from the harmonic analysis.}

\noindent\begin{tabular}{rrrlr}
\hline \\
$j$      & $A_j^c $     & $A_j^s $    &$\sigma^2_j$          &$P_j$\\
\hline \\
 1       &  $  5.634826$& $-46.439608$ &   $.20    $            &$222.28$\\
 2       &  $  -.531603$& $   .283347$ &   $.027   $            &$ 19.54$\\
 3       &  $  -.051990$& $   .172145$ &   $.011   $            &$ 12.54$\\
 4       &  $  -.001363$& $  -.117686$ &   $.0039  $            &$ 46.04$\\
 5       &  $  -.037362$& $   .009689$ &   $.0021  $            &$ 22.22$\\
 6       &  $  -.029915$& $   .004394$ &   $.0016  $            &$  5.84$\\
 7       &  $  -.017975$& $   .018877$ &   $.0013  $            &$  7.90$\\
 8       &  $  -.002987$& $  -.020452$ &   $.0011  $            &$  1.00$\\
 9       &  $  -.007429$& $   .017729$ &   $.00090 $            &$  6.50$\\
10       &  $  -.004306$& $   .016504$ &   $.00076 $            &$  4.63$\\
11       &  $  -.000759$& $   .016891$ &   $.00062 $            &$  9.26$\\
12       &  $   .002933$& $  -.013173$ &   $.00052 $            &$  3.57$\\
13       &  $  -.010059$& $  -.006987$ &   $.00045 $            &$  5.28$\\
14       &  $  -.001567$& $  -.010152$ &   $.00040 $            &$  4.08$\\
15       &  $   .006573$& $   .007869$ &   $.00034 $            &$   .50$\\
16       &  $  -.002887$& $   .004019$ &   $.00033 $            &$  1.09$\\
17       &  $   .000452$& $  -.000316$ &   $.00033 $            &$   .17$\\
\hline
\end{tabular}\vglue.2cm
\vskip.2cm
\vskip.3cm
\noindent\includegraphics[width=8cm,height=6cm]{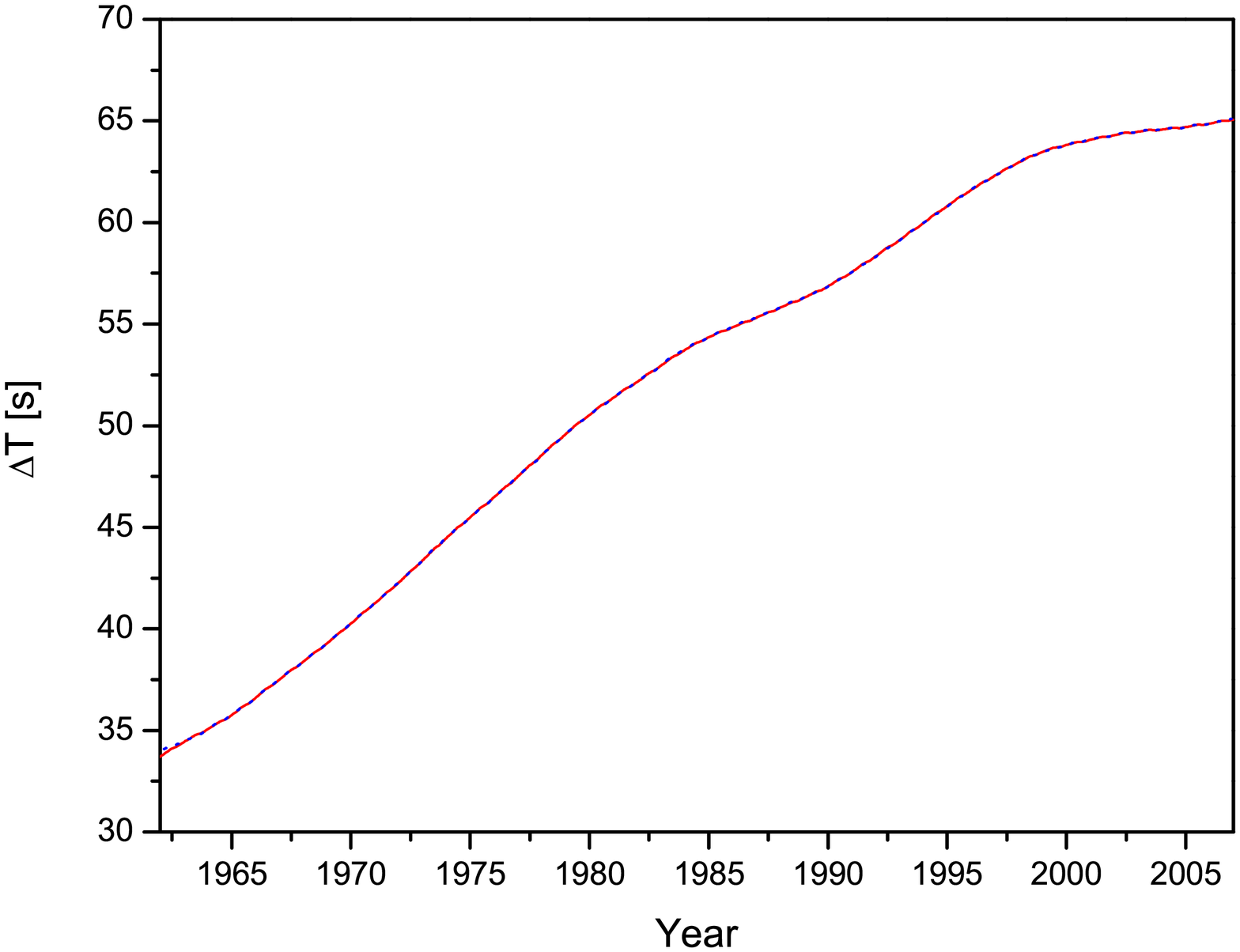}

\noindent{\bf Figure 2.} {\sl The dashed line (blue) represents IERS
data and the solid one (red) represents our approximation
(prediction) between  1962.15 and 2006.95. }

The coefficients of the linear function (in equation (1)) are: free
term $C_1=3.408457$ and slope coefficient  $C_2=0.357042$. The
coefficients of the harmonic components are given in Table 1.
$\sigma^2_j$ is the upper limit for the coefficients of the harmonic
components (it is the same for both coefficients, $A_j^c $  and
$A_j^s $) and it is given with two significant digits. The periods
$P_j$ are given in years. Because of the secular and decade
variations all terms corresponding to insignificant periods and
amplitudes (smaller than few milliseconds) are on the noise level so
that they are not determined.


In Fig. 2 the dashed line (blue) represents the IERS data from
1962.15 to 2006.95, whereas the solid one (red) represents our
approximation for the period between  1962.0 and 2007.0. One can
notice an excellent agreement.


According to our approximation till 2024 there is practically  no
need to introduce leap seconds because an accumulation of 0.9 seconds
is reached as late as at 2014.0, i.\,e. an accumulation of one second
in 2015. The local  maximum in our approximation occurs at 2016.4 and
its value is 65.944 which exceeds the difference of one second by 0.1
seconds only (Fig. 3). In view of the further negative trend of our
approximation the next leap second (negative) should be introduced in
2024 if the difference of 0.1 seconds at the moment of approximation
maximum were neglected.


\vskip.2cm

\noindent\includegraphics[width=8cm,height=6cm]{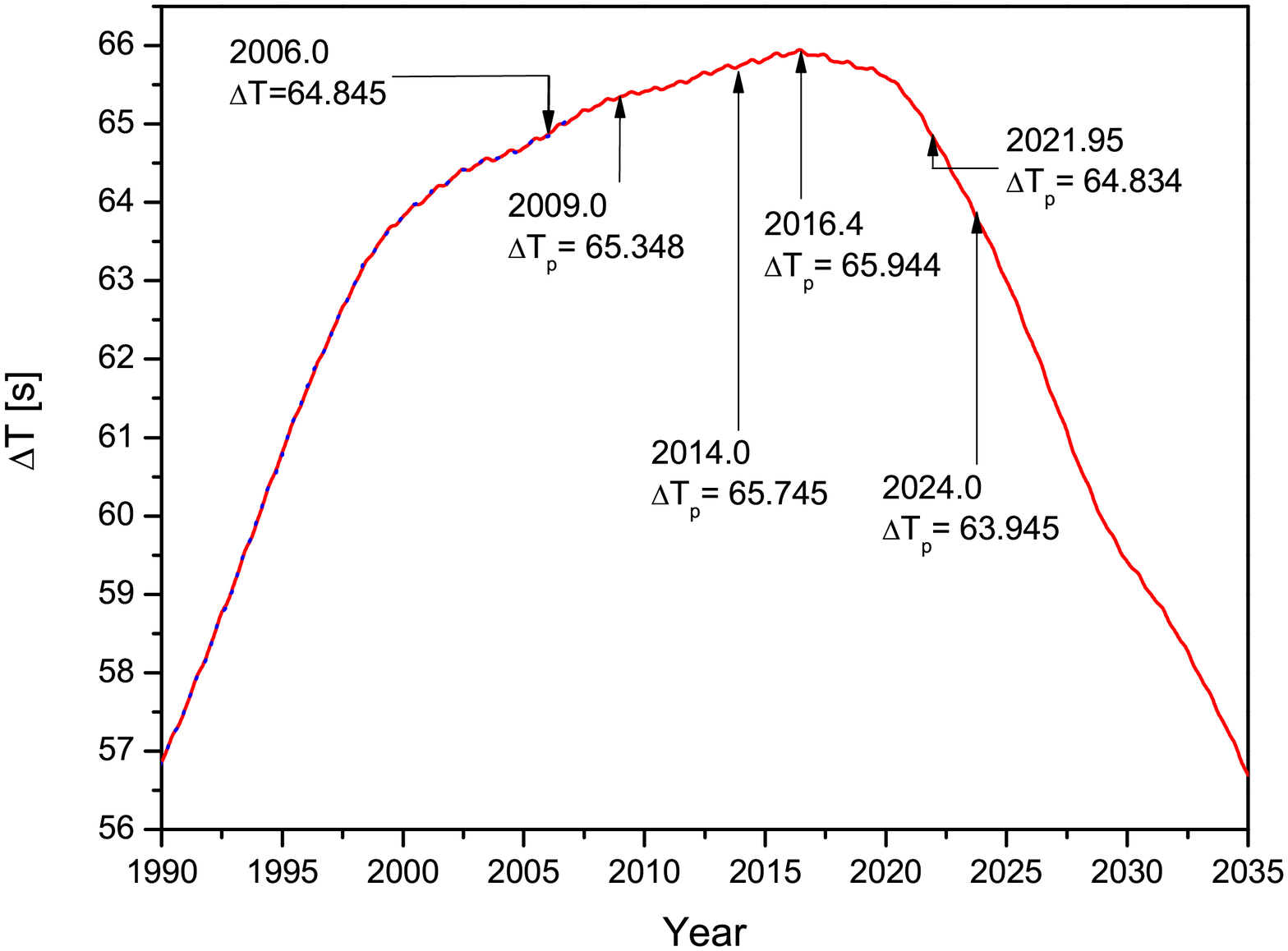}

\noindent{\bf Figure 3.} {\sl The approximation from 1990 to 2035 and
IERS data till 2006.95. The indicated points are: the moment when the
leap second was introduced for the last time (2006), the next
introducing (2009), the point at which according to our approximation
the  difference of 0.9 seconds between $UT1$ and $UTC$ is reached
(2014.0), the local maximum of our approximation (2016.4), the point
for which  $\Delta T$  is approximately the same as in 2006 when the
leap second was introduced for the last time (2021.95) and the point
when, according to our approximation, the next leap second (negative)
should be introduced (2024.0). Note that this means to introduce only
one negative leap second within 18 years (2006--2024).  }


\vskip.2cm

\noindent\includegraphics[width=8cm,height=6cm]{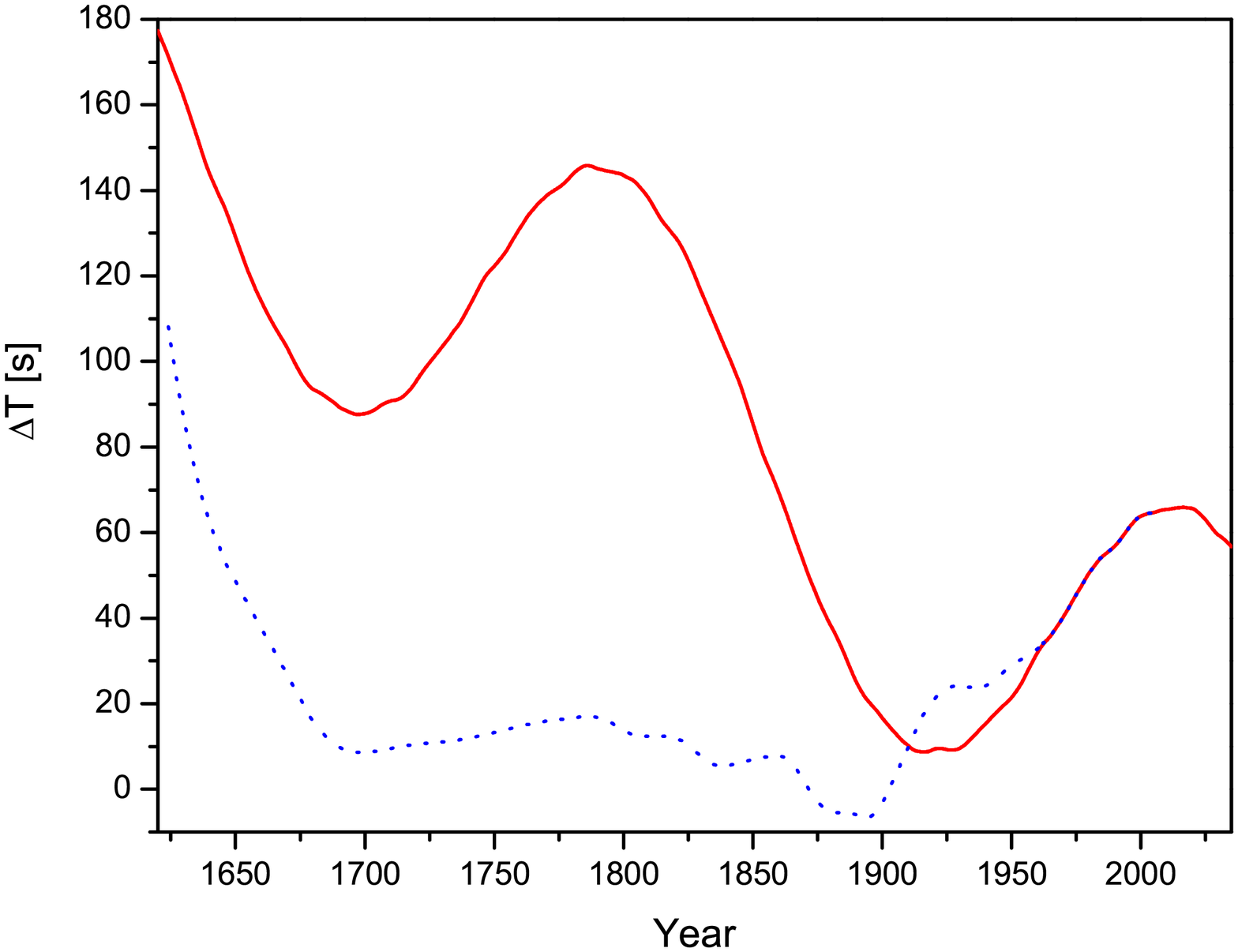}

\noindent{\bf Figure 4.} {\sl The historical observations covering
the period  1620--2006.95 are represented by the dashed line (blue)
and our approximation concerning the interval 1620--2035 by the solid
line (red). }


\vskip.2cm

\section{4. CONCLUSIONS}

As the day at present is actually closer to 86400 SI seconds the leap
seconds have not been required regularly. However, this condition
cannot persist and the long or mid-term trend will be eventually
restored. The asymptotic behaviour of the $\Delta T$ in the
Stephenson--Morrison approximation (1995) is not natural.

As can be seen from the analysis (Fig. 4) applied within a relatively
short time interval, 1962.15--2006.95, the extrapolation of the
$\Delta T$ value agrees well with both the historical values
(1620-1962) and the  real measurements from 1962.15 to 2006.95. The
discrepancy in the amplitude is higher for the former period, but the
agreement for the phase is very good which indicates that the
approximation could be improved if a  bigger  set of measurements
were available.

Our conclusion, that  the first (additional) leap second should be
introduced  in 2014 only, opposes to all the extrapolations proposed
till nowadays by both international institutions and individual
experts. The majority of them predicts 2008 as the year when the
additional correction should be introduced.

The values attained by $\Delta T$ are of such order that all the
physical factors unambiguously recognized till nowadays cannot cause
such a phenomenon and in this sense the present analysis means  a
strictly  mathematical approximation only. We believe that the years
to come will show the correctness of the obtained results.




\bigskip


\references

\def\ee{\  }



%

%


Eubanks, T.M., 1993, Variations in the orientation of the Earth. In:
Smith, D. E., Turcotte, D.L. (Eds.), Contributions of Space Geodesy
to Geodynamics: Earth Dynamics. Geodyn. Ser. vol. 24, pp. 1–54.

%

%

%
%
%

%
%


 Lambeck K.\ 1980\ {\it The Earth's Variable Rotation: Geophysical
Causes and Consequences}\ (Cambridge Univ. Press).\ %

McCarthy, D. D. (ed.), 1996, IERS Conventions, IERS Technical Note
21, Observatoire de Paris, Paris.

Munk W. H., MacDonald G. J. F.\ 1960\ {\it The Rotation of the
Earth: A Geophysical Discussion}\ (Cambridge Univ. Press).\ %

%


Salstein, D.A. and R.D. Rosen, 1989: Regional contributions to the
 atmospheric excitation of rapid polar motions. J. Geophys. Res., 94, {\bf 9971}--9978.

Stephenson,  F. R., and Morrison L. V., 1995, Long-Term Fluctuations
in the Earth's Rotation: 700 BC to AD 1990, {\it Phil. Trans. R.
Soc. Lond.}, Vol. 351, No. 1695., pp. {\bf 165}--202.

%

{{\SH}egan, S.}, Damjanov, I. and {\SH}urlan, B., 2003, Earth's
rotation irregularities derived from UTI$_{BLI}$ by method of
multi-composing of ordinates, {\it Serbian Astronomical Journal},
Vol. 167, {\bf 53}.

%
%
%

Vondr\'ak J., 1987 in Holota P. (Ed.) {\it Proc. Internat. Symp.
 Figure and Dynamics of the Earth, Moon and Planets}\ (Astron. Inst. and Res.
Inst. Geod. Praha), {\bf 1039}.\ %

%


Vondr{\'a}k, J., Pe{\sh}ek, I., Ron, C., {\CH}epek, A.: 1998, Earth
orientation parameters 1899.7--1992.0 in the ICRS based on the
HIPPARCOS reference frame, {\it Astron. Inst. of the Academy of
Sciences of the Czech R.}, Publ. No. 87.

%

%



Zharkov, V.N., Molodensky, S.M., Brzezinski, A., Groten, E., Varga,
P., 1996, The Earth and its rotation. Wichmann, Heidelberg, vol.
XIII, {\bf 501}.

\endreferences
}

\end{multicols}
\vglue-2cm


\naslov{TEXKO{\CJ}E U OQUVANJU PRESTUPNE SEKUNDE}


\autori{S. Xegan$^{1}$, B. Xurlan$^{1}$, S. Vidojevi{\cj}$^1$}

\vskip3mm


\adresa{$^1$Katedra za astronomiju, Matematiqki fakultet, Univerzitet
u Beogradu\break Student\-ski trg 16, 11000 Beograd, Srbija}

\vskip.7cm


\centerline{UDK \udc}


\centerline{\rit ...........................}

\vskip.7cm

\begin{multicols}{2}
{


\rrm Ispitana je mogu{\cj}nost da se prognozira broj prestupnih
sekundi u bliskoj budu{\cj}nosti na osnovi poznavanja nekoliko
harmonijskih komponenti u Zemljinoj rotaciji. Anali\-zi\-ra\-ni su
podaci {\rm IERS}-a koji obuhvataju interval od 1962,15 do 2006,95.
Razlika $\Delta T$ je ekstrapolirana do 2035. godine i uporedjena sa
predvidjanjima {\rm IERS}-a do 2012. godine. Mozhe se videti da je u
intervalu od 2006. do 2024. godine potrebno uvesti samo jednu
(negativnu) prestupnu sekundu.}
\end{multicols}

\end{document}